\documentclass{article}
            \usepackage{graphicx}
            \usepackage{amsmath,amssymb}
\begin{document}
            \title{Effects of a scalar scaling field on quantum mechanics.}
            \author{Paul Benioff,\\
            Physics Division, Argonne National
            Laboratory,\\ Argonne, IL 60439, USA \\
            e-mail:pbenioff@anl.gov}

            \maketitle


            \begin{abstract}
            This paper describes the effects of a complex scalar scaling field on quantum mechanics. The field origin is an extension of the gauge freedom for basis choice in gauge theories to the underlying scalar field. The extension is based on the idea that the value of a number at one space time point does not determine the value at another point. This, combined with the description of mathematical systems as structures of different types, results in the presence of separate number fields and vector spaces as  structures, at different space time locations. Complex number structures  and vector spaces at each location, are scaled by a complex space time dependent scaling factor. The effect of this scaling factor on several physical and geometric quantities has been described in other work.  Here the emphasis is on quantum mechanics of one and two particles, their  states and properties. Multiparticle states are also briefly described.  The effect shows as a complex, nonunitary,  scalar field connection on a fiber bundle description of  nonrelativistic quantum mechanics.  The lack of physical evidence for the presence of this field so far means that the coupling constant of this field to fermions is very small. It also means that the gradient of the field must be  very small in a local region of cosmological space and time.  Outside this region there are no restrictions on the field gradient.
            \end{abstract}

            \section{Introduction}
            The  relationship between mathematics and physics at a basic level is of much interest to many. It was most clearly raised some time ago by Wigner in a paper on "The unreasonable effectiveness of mathematics in the natural sciences" \cite{Wigner}. This work resulted in much discussion  about the relationship  \cite{Omnes,Plotnitsky}.  There was also the suggestion that physics is mathematics \cite{Tegmark}.

            These ideas led to work towards a coherent theory of mathematics and physics together in which physics and mathematics are treated together as a coherent whole \cite{BenCTPM,BenCTPM2}. The origin of continuing work on this topic \cite{BenNOVA,BenQSMF} was based on gauge theory considerations. The mathematical setup for these theories consists of vector spaces associated with each point of a space time manifold. Unitary connections between these spaces express the concept,  based on work by Yang and Mills \cite{Yang}, of the freedom of choice of bases in vector spaces.

             This work is based on an extension of the freedom of basis choice in vector spaces to the freedom of number value choice for  the underlying  scalar fields. The choice freedom is based on the observation that the value of a number at one space time point does not determine the value of the same number at another point. This results in the extension of gauge theory to include separate scalar fields associated with the vector spaces at each space time point.

             This localization of scalar fields follows from the representation of mathematical systems, such as vector spaces and scalar fields, as mathematical structures and relations between the structures \cite{Shapiro,Barwise,Keisler}.  Structures consist of a base set, a few basic operations and relations, and none or a few constants. They satisfy axioms relevant to the system type being considered. It can be shown that, for each number type, there exist many  different structures that differ from one another by scaling factors. It follows that base set elements, as numbers, do not have intrinsic, structure independent values.  They  have different values in different scaled structures. This fact and the choice freedom of number values at different locations results in the existence of separate local scaled number structures at each point.

            The scaling used here is linear in that the sum of two scaled numbers equals the scaling of the sum of the numbers. This type of scaling has recently been generalized to nonlinear scaling of number structures \cite{Czachor}.  The additional complexity of this type of scaling makes it difficult to apply it to different areas of physics.  These applications are work for the future.

            Relations between number structures at different locations are implemented by a scalar scaling field. The value of the field at each location is the scaling factor for the number structure at each point. The field is not unitary as it is the product of a real factor and a phase factor.

            In earlier work extension of gauge theory to include the effect of the scaling scalar field was described.  The effect of the scaling field and use of local scaled number structures was extended to briefly describe some other physical and geometric quantities \cite{BenNOVA,BenQSMF}.  In this work  the emphasis is on quantum mechanical quantities for one, two, and multiparticle systems. The definition of the scaling field is extended to apply to entangled states of two or more particles.

            Fiber bundles \cite{Husemoller,Husemoller2} have been much used in physics. They have been used to describe gauge theories \cite{Daniel,Drechsler} and   nonrelativistic quantum mechanics \cite{Moylan,Sen,Bernstein,Iliev,Asorey}. The quantum descriptions include an introductory description \cite{Bernstein}, a detailed mathematical development \cite{Iliev} and descriptions in which symmetry groups and semigroups  serve as the base space \cite{Sen,Asorey}.

            This paper uses fiber bundles to describe the effect of a scaling scalar field on one, two, and multiparticle  quantum systems in  nonrelativistic quantum mechanics. The description includes the  effect of the field on multiparticle spatially entangled states.  The plan of the  paper is to first describe the scaling of scalar fields and vector spaces.  This is done in the next section.  Local representations of these and other structures at each point of a space or space time manifold are conveniently described by use of fiber bundles.  These are described in  Section \ref{FB}. The bundle fibers are large in that they contain sufficient scaled mathematical structures to describe some aspects of quantum mechanics. In earlier work that was not based on fiber bundles \cite{BenNOVA,BenINTECH},  mathematical universes or the mathematics available to observers were the equivalent of fibers at different locations.

            The scalar scaling field is introduced in the next section on connections. The field is complex valued since it acts on Hilbert spaces and on complex numbers as a scalar base for the Hilbert spaces. Section \ref{QM} describes some aspects of the use of fiber bundles in quantum mechanics. The three subsections discuss one, two, and multiparticle  particle states. The discussion includes entangled two and multiparticle states. The fiber bundles are used to describe a projection of a wave packet onto the bundle fibers. This is followed by the use of the scaling field connections to map the projected wave packet to a fiber at an arbitrary reference location. Without the localization, the implied space integration of the wave packet amplitudes is undefined. An expansion of the definition of the connection for single particle states to accommodate two and multiparticle states is described.

            The discussion section \ref{D} concludes the paper. Some justification for the localization of quantum mechanics is given.  Also coupling and cosmological limitations on the  scalar scaling field are noted.

            \section{Scaling of scalar and vector space structures}\label{SVSS}
            One begins with the observation that mathematical systems of different types can be represented as different structures \cite{Shapiro,Barwise,Keisler}. A structure  consists of a base set of elements and a few basic operations, relations, and constants. The structure is required to satisfy a set of axioms relevant to the structure type being considered.  In addition there are maps and operations between structures of different types.  Examples of such maps are scalar vector multiplication and norms of vectors in vector spaces.

            \subsection{Number structures}
            Number structures of different types are important examples of scalar structures.  These structures can all be scaled.  The types include the natural numbers, integers, and rational, real, and complex numbers.  In addition mathematical structures that include numbers in their descriptions can also be scaled.  This includes vector spaces, operator  algebras, and group representations.  Here the discussion will be limited to complex numbers and vector  spaces with complex numbers as the associated scalars.

            A structure for the complex number field is \begin{equation}\label{bC}
            \bar{C}=\{C,+,-,\times,\div,^{*}, 0,1\}.\end{equation} This structure satisfies the axioms for an
            algebraically closed field of characteristic $0$ plus
            axioms for complex conjugation \cite{complex}.

            In order to describe scaling it is important  to distinguish the base set elements of a structure from the values or meaning they have in a structure.  By themselves, base set elements have no intrinsic value.  The values they get depend on the structure containing them.  Here many different complex number structures  that differ by scaling factors will be considered. It follows that a value of a  base set element  will depend on the scaling factor for the structure containing it.

            To describe this in more detail, it is useful to distinguish two types of structures: a value structure and different representation structures. For complex numbers the structure $\bar{C}$ of Eq. \ref{bC} represents a complex number value structure. The  abstract elements of the base set , $C$, and  constants, $0$ and $1$ have meaning as number values. The  four field operations and complex conjugation  also have meaning. Elements of $C$ will always be referred to as  complex number values. This distinguishes them from elements of the base set of representation structures.  They are referred to as complex numbers. Also structures are distinguished from base sets by an overline. For example $C$ is the base set of the structure $\bar{C}.$

            The many complex number representation structures are distinguished from the value structures by the scaling factor used as a subscript.  For  complex scaling factors, $s$ and $t,$ the associated scaled structures, $\bar{C}^{s}$ and $\bar{C}^{t}$ are given by\begin{equation}\label{bCts} \begin{array}{c}\bar{C}^{t}= \{C_{t},\pm_{t}, \times_{t},\div_{t},(-)^{*_{t}},0_{t}, 1_{t}\}\\\\\bar{C}^{s}=\{C_{s},\pm_{s},\times_{s}, \div_{s},(-)^{*_{s}},0_{s}, 1_{s}\}.\end{array}\end{equation}Here $C_{t}$ and $C_{s}$ are the same sets of numbers.  However the value of a given element of $C_{t}$ is different than its value in $C_{s}$ because it belongs to a different complex number structure.

             Numbers in the base set can be represented in the form $a_{t}$.  This represents the number in $C_{t}$ in $\bar{C}^{t}$ that has value $a$ in $\bar{C}.$ $0_{t},1_{t}$ and $0_{s},1_{s}$ are examples of this.

            Value maps of the components of representation structures onto the value structure, $\bar{C},$ illustrate some properties of numbers and their values.  Let $a_{s}$ and $a_{t}$ be numbers in $C_{s}=C_{t}.$ Let $v_{s}$ and $v_{t}$ be value maps of $\bar{C}^{s}$ and $\bar{C}^{t}$ onto $\bar{C}.$ The two numbers, $a_{s}$ and $a_{t}$, have the same value, $a$. However they are different numbers.  This can be seen from\begin{equation}\label{vtat}v_{t}(a_{t})=a=v_{s}(a_{s}) \end{equation} and\begin{equation}\label{vsat}v_{s}(a_{t})=\frac{t}{s}v_{s}(a_{s}) =\frac{t}{s}v_{t} (a_{t})=\frac{t}{s}a.\end{equation}This equation holds of $s$ and $t$ are exchanged.

\begin{figure}[h]\begin{center}\vspace{2cm}
           \rotatebox{270}{\resizebox{150pt}{150pt}{\includegraphics[170pt,200pt]
            [520pt,550pt]{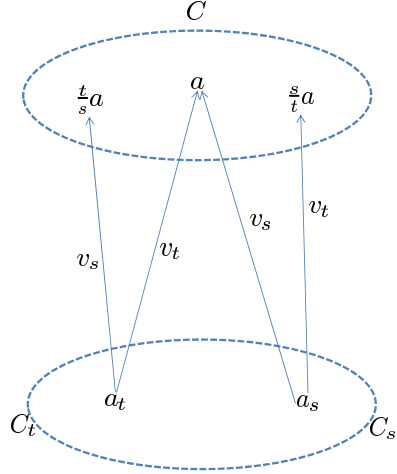}}}\end{center}\caption{Illustration of the effect of complex number structure membership on the values of base set elements. The numbers, $a_{s}$ and $a_{t}$, as respective members of  $\bar{C}^{s}$ and $\bar{C}^{t}$, have the same value, $a$.  Yet they are different numbers. The fact that $C_{s}$ is the same base set of numbers as $C_{t}$ is also shown.}\label{NSQM1}\end{figure}

            The relations between the base set numbers, $a_{s},$ and $a_{t}$ and their values in different structures are shown in Figure \ref{NSQM1}.  The figure shows clearly that $a_{s}$ and $a_{t}$ are different numbers even though they have the same value in their respective structures.

            The action of $v_{t}$ on the components of $\bar{C}^{t}$ is given by \begin{equation}\label{vtCt}
            \begin{array}{c}v_{t}(a_{t})=a,\;\;v_{t}(+_{t})=+,\\\\ v_{t}(\times_{t})=\times,\;\;v_{t}(-)^{*_{t}} =(-)^{*},\\\\ v_{t}(0_{t})=0,\;\;v_{t}(1_{t})=1.\end{array}\end{equation}This definition holds for $v_{s}$ and $\bar{C}^{s}$  with $t$ replaced everywhere by $s.$

            This definition shows that the value functions are all bijections.  As such they have inverses.  For each $s$, $v^{-1}_{s}$ where $v^{-1}_{s}\bar{C}=\bar{C}^{s}.$ This definition is closer to that in \cite{Czachor}.

            So far valuations of the different number structures, such as $\bar{C}^{t}$ and $\bar{C}^{s}$ have been described. What is needed is a relative valuation of the components of one structure as a function of the components  of another structure.   This will be much used later on.

            This is achieved by a map of the components of $\bar{C}^{t}$ onto those of $\bar{C}^{s}$ such that Eq. \ref{vsat} is satisfied.  Let $Z^{t}_{s}$ be such a map.  The action of $Z^{t}_{s}$ on the components of $\bar{C}^{t}$ is given by \begin{equation}\label{Zts}
            \begin{array}{c}Z^{t}_{s}(a_{t})=\frac{\mbox{\small $t_{s}$}}{\mbox{\small $s_{s}$}}a_{s},\;\;Z^{t}_{s}(+_{t})=+_{s},\\\\ Z^{t}_{s}(\times_{t})=\frac{\mbox{\small $s_{s}$}}{\mbox{\small $t_{s}$}}\times_{s},\;\;Z^{t}_{s}(a_{t}^{*_{t}}) =\frac{\mbox{\small $t_{s}$}}{\mbox{\small $s_{s}$}}(a_{s}^{*_{s}}),\\\\ Z^{t}_{s}(0_{t})=0,\;\;Z^{t}_{s}(1_{t}) =\frac{\mbox{\small $t_{s}$}}{\mbox{\small $s_{s}$}}1_{s}.\end{array}\end{equation}

            This definition shows that $Z^{t}_{s}$ defines a number structure, $\bar{C}^{t}_{s},$ that represents the components of $C^{t}$ in terms of those of $\bar{C}^{s}.$  $\bar{C}^{t}_{s}$ is given by\begin{equation}\label{BCts}Z^{t}_{s}\bar{C}^{t}=\bar{C}^{t}_{s}= \{C_{s},+_{s}, \frac{s_{s}}{t_{s}}\times_{s},\frac{t_{s}}{s_{s}}(a_{s}^{*_{s}}),0_{s}, \frac{t_{s}}{s_{s}}1_{s}\}. \end{equation}Here $t_{s}$ and $s_{s}$ are the numbers in $C_{s}$ that have $s$ values, $t$ and $s,$ in $\bar{C}.$

            Note that $Z^{t}_{s}$ is the identity map on the base set.  This follows from Eq. \ref{vsat}
            \begin{equation}\label{vsatv}v_{s}(a_{t})=v_{s}(\frac{t_{s}}{s_{s}}a_{s})=\frac{t}{s}v_{s}(a_{s}) =\frac{t}{s}v_{t}(a_{t})=\frac{t}{s}a.\end{equation} and the fact that $v_{s}$ is a bijection. These results show that $a_{t}=(t_{s}/s_{s})a_{s}.$

            The description of $\bar{C}^{t}_{s}$  in Eq. \ref{BCts} seems strange because $(t_{s}/s_{s})1_{s}$ can have a complex value and satisfy the multiplicative identity axiom.  This follows from $v_{s}(t_{s}/s_{s}1_{s})=t/s$ and $$\frac{t_{s}}{s_{s}}1_{s}(\frac{s_{s}}{t_{s}} \times_{s})\frac{t_{s}}{s_{s}}a_{s} =\frac{t_{s}}{s_{s}}a_{s}.$$ The properties of complex conjugation are also satisfied in that $$(\frac{t_{s}}{s_{s}}a_{s}(\frac{s_{s}}{t_{s}} \times_{s})\frac{t_{s}}{s_{s}}b_{s})^{*_{s}}= (\frac{t_{s}}{s_{s}})^{*_{s}}(a_{s})^{*_{s}}) (\frac{s_{s}}{t_{s}})^{*_{s}}\times_{s}(\frac{t_{s}}{s_{s}})^{*_{s}})(b_{s}^{*_{s}}) =(\frac{t_{s}}{s_{s}})^{*_{s}}(a_{s}^{*_{s}}\times_{s}b^{*_{s}}_{s}).$$ Note one must include here the complex conjugation of the factor multiplying the multiplication operator. Also one has $$v_{s}((\frac{t_{s}}{s_{s}} a_{s})^{*_{s}})^{*_{s}}=(v_{s}(\frac{t_{s}}{s_{s}}a_{s})^{*_{s}})^{*} =((\frac{t}{s}a)^{*})^{*} =\frac{t}{s}a.$$ It follows from this and the bijective property of $v_{s}$ that $$((\frac{t_{s}}{s_{s}}a_{s})^{*_{s}})^{*_{s}}=\frac{t_{s}}{s_{s}}a_{s}.$$

            Some specific examples will help to make  clearer some of the properties of the scaled complex number structures.   The first example is for complex numbers that are also real rational numbers. Values of these  numbers will have the usual decimal form, such as $12.47.$   Let  $t=43.1$ and $s=1.67.$ The numbers, $12.47_{43.1}$ in $\bar{C}^{43.1}$ and $12.47_{1.67}$ in $\bar{C}^{1.67}$ have the same value, $12.47$, in $\bar{C}.$ However, $12.47_{43.1}$ is a different number in $C_{43.1}=C_{1.67}$ than $12.43_{1.67}.$

             The same results hold for complex rational numbers.  For example, let $t=-.006+4.1i$ and $s=50.2-6.4i$ and $a=-27.1+3.7i.$ Then $a_{s}$ and $a_{t}$  have the same respective $s$ and $t$ value, $a$ in $\bar{C}.$ But they are different numbers. These results extend to more general complex numbers.  For example let $t=\sqrt{2}-i\sqrt{4.7}$ and $s=\pi+ie$ and $a=\sqrt{5}+i\sqrt{7}.$ Then $a_{s}$ and $a_{t}$ have the same $s$ and $t$ value, $a=\sqrt{5}+i\sqrt{7}.$ But they are different numbers in $C_{t}=C_{s}.$

            The above description of the effects of number scaling  uses a representation of numbers as $a_{t}$ where $a$ is the value of the number in $\bar{C}^{t}$.  This representation is rather abstract as it makes no connection between  numbers and other types of number  representations that are more in tune with the usual usage.

            An example is the representation of numbers based on finite strings over an alphabet, $a,b,\cdots,j$ with $a<b<\cdots,<j$. Add two elements, $+,-,$ and a point to the alphabet.  These strings with lexicographic linear ordering, represent a big subset of the rational numbers in the fixed point representation. Examples of strings ar $dfa.ggi$ and $a.aafgdh.$

            In order to assign rational number values to these strings, two choices are needed: the element to have the value $0$ and the element to have the value $1.$ Let the string $a.aa$ and equivalents\footnote{Equivalents means strings with an arbitrary number of $a's$ to the right end of the string.} have the value $0.$

            The usual choice for the string  to have the value $1$ is $b.a$ and equivalents. However any other choice is possible. The choice determines the scaling factor.  For example, let the string $dbf.aag$ have the value $1.$ Then the scaling factor, $t,$ is the value of $dbf.aag$ in the usual structure.  This value is $215.006.$

            The strings can be used to define scaled complex number structures where the scaling factor is rational. For example the string $dbf.aag$ is the number $1_{t}$ in the structure $\bar{C}^{t}$. Also $v_{t}(dbf.aag)=1.$

            The value of any other string, such as $-a.jjhgbi$ in $\bar{C}^{t},$ is given by $$v_{t}(-a.jjhgbi)=\frac{1}{t}v_{1}(-a.jjhgbi).$$ $\bar{C}^{1}$ is the usual structure in which $b.a$ has the value $1.$ In this structure, $v_{1}(dbf.aag)=215.006.$  From this one has, $$v_{t}(-a.jjhgbi)=\frac{-0.997618}{215.006}.$$ Also $v_{t}(b.a)=1/215.006.$

            These representations show that an individual string has no intrinsic value.  Its value depends on the structure containing it.  For example,  the string, $b.a$ has value $1$ in $\bar{C}^{1},$ and value $1/215.006$ in $\bar{C}^{t}$. This scaling dependence extends to real numbers as infinite alphabet strings or equivalence classes of Cauchy \cite{Cauchy} sequences of Dedekind cuts of rational number strings, and to complex numbers as pairs of real numbers. If $\psi$ and $\phi$ are infinite alphabet strings or Cauchy sequences of rational numbers and $t$ and $s$ are complex number values, then the value of $\psi+_{t}i\phi$ in $\bar{C}^{t}$, relative to that in $\bar{C}^{s}$ is given by  $$v_{t}(\psi+_{t}i\phi)=\frac{s}{t}v_{s}(\psi+_{s}i\phi).$$

            \subsection{Vector spaces}\label{VS}

            As noted before,  number scaling affects types of mathematical structures whose description includes maps to scalars,  Vector spaces with norms or scalar products and  vector scalar multiplication are examples of  such structures. Because of these maps, vector space structures should not be considered in isolation.  Instead the associated scalar structure should  be included.

            As an example let $\bar{C}\times\bar{V}$ be a combined scalar vector space structure.  If $\bar{V}$ is a normed vector space, then \begin{equation}\label{BV}\bar{V}=\{V,\pm,\cdot, |f|, f\}\end{equation} Here $|f|$ denotes the real valued norm in $\bar{C},$  the dot denotes multiplication of vectors in $\bar{V}$ by scalar values in $\bar{C}$, and $f$ is an arbitrary vector.  Hilbert spaces, with $|f|$ replaced by $\langle g|f\rangle,$ are  examples.  Tangent spaces, over a geometric manifold are  associated with real scalars.

            Number scaling affects both $\bar{C}$ and $\bar{V}$. For each scaling factor, $s$, the corresponding scalar vector space structure pair is $\bar{C}^{s}\times\bar{V}^{s}$ where \begin{equation} \label{BVs}\bar{V}^{s}=\{V_{s},\pm_{s}, \cdot_{s},|f_{s}|_{s},f_{s}\}.\end{equation} Here the distinction between number and number value structures is applied to vector spaces. $\bar{V}$ is assumed to be a vector value structure and structures, $\bar{V}^{s}$, are vector structures. Vectors in $V_{s}$ are represented by $f_{s}$ where $f$ is the vector value for $f_{s}.$  That is $v_{s}(f_{s})=f.$

            If $\bar{C}^{t}_{s}$ is the  projection or relative scaling of $\bar{C}^{t}$ onto $\bar{C}^{s},$ as in Eq. \ref{BCts}, then the corresponding scalar vector space structure pair is $\bar{C}^{t}_{s}\times\bar{V}^{t}_{s}$ where\footnote{There is another representation of $\bar{V}^{t}_{s}$ in which the vectors do not scale.  This is $\bar{V}^{t}_{s}=\{V,\pm_{s},(s/t)\cdot_{s},(t/s)|f_{s}|_{s}, f_{s}\}.$  This is not used here because the equivalence between $n$ dimensional vector spaces and $\bar{C}^{n}$ (or $\bar{R}^{n}$) fails for this $\bar{V}^{t}_{s}$.   $\bar{V}^{t}_{s}$ is not equivalent to $(\bar{C}^{n})^{t}_{s}$ or to $(\bar{R}^{n})^{t}_{s}.$)}\begin{equation}\label{BVts} \bar{V}^{t}_{s} =\{V_{s},\pm_{s},\frac{s_{s}}{t_{s}}\cdot_{s},\frac{t_{s}}{s_{s}}|f_{s}|_{s}, \frac{t_{s}}{s_{s}}f_{s}\}.\end{equation}

            As was the case for the scalar structures, the scaling of the components of $\bar{V}^{t}_{s}$ is not arbitrary. It is done so that the validity of the relevant axioms is preserved under scaling.  The definition of value maps, used for complex numbers, can be applied to vector space components. Examples are $$v_{s}(a_{t}\cdot_{t}f_{t})=v_{s}(\frac{t_{s}}{s_{s}}a_{s}\frac{s_{s}} {t_{s}}\cdot_{s}\frac{t_{s}}{s_{s}}f_{s})=v_{s}(\frac{t_{s}}{s_{s}}(a_{s}\cdot_{s}f_{s})) =\frac{t}{s}(a\cdot f)$$ and $$v_{s}(|f_{t}|_{t})=v_{s}(\frac{t_{s}}{s_{s}}|f_{s}|_{s})= \frac{t}{s}|f|.$$  Note that both  $|f_{t}|_{t}$ and $|f_{s}|_{s}$ have the same respective $v_{t}$ and $v_{s}$ value, $|f|,$ in $\bar{C}$.  But $v_{s}(|f_{t}|_{t})$ real if and only if $t/s$ is real. For scalar products in Hilbert spaces one has $$v_{s}(\langle f_{t},g_{t}\rangle_{t})=\frac{t}{s} v_{t}(\langle f_{t},g_{t}\rangle_{t})=\frac{t}{s}\langle v_{t}(f_{t}),v_{t}(g_{t})\rangle =\frac{t}{s}\langle f,g\rangle.$$

            \section{Fiber Bundles}\label{FB}

            As was noted in the introduction, fiber bundles have been used as a framework to describe quantum systems in nonrelativistic quantum theory \cite{Moylan}-\cite{Asorey}. Here they are used to describe the effects of a scalar scaling field on some simple properties of quantum systems. The use of Euclidean space as the base space of the bundles, and local mathematical systems at each base space point, enables the local description of nonlocal properties, such as those described by space integrals and derivatives (as local limits of nonlocal properties), of quantum systems.

            A fiber bundle \cite{Husemoller,Husemoller2} is a triple, $E,\pi,M$ where $E$ is the total space, $\pi$ is a projection of $E$ onto $M$ and $M$ is the base space.  The inverse map, $\pi^{-1}$, maps points of $M$ onto fibers in $E$.  Since $M$ is a flat space in this work, the fiber bundle  is a product bundle, $M\times F,\pi,M$.  $F$ is the fiber and $\pi^{-1}(x)$ is  a copy of $F$  at $x.$

            Here earlier work \cite{BenQSMF,BenSPIE5} on including scalars and other mathematical structures with vector spaces in the fibers is used. Justification for  this localization is based on an extension of the localization of vector spaces used in gauge theories to scalar fields.

            The argument for localization of scalar fields, such as complex numbers, is similar to that of Yang Mills \cite{Yang} for localization of vector spaces. The  position taken here is that   the value of a number at one point of space time does not determine the value of the same number at another point. This position, combined with the observation that the  values or meaning of the base set numbers  are determined by the structure containing them, leads to separate scalar field structures at each space time point.

            The structures are distinguished by scaling factors.  For complex number structures, the scaling factors are complex. The fiber bundle for the scaled complex numbers and vector spaces has the fiber, $F,$ defined by \begin{equation} \label{F}F=\bigcup_{s}(\bar{C}^{s}\times\bar{V}^{s}).\end{equation} Here $\bar{V}^{s}$ is a scaled vector space with $\bar{C}^{s}$ the associated  scaled scalar field. The union is over all scaling factors, $s,$ as values in $\bar{C}.$ The fiber at $x$ is defined by\begin{equation}\label{pim1x}\pi^{-1}(x)=\bigcup_{s}(\bar{C}^{s}_{x} \times\bar{V}^{s}_{x}).\end{equation} The mathematical structures in the fiber at $x$ are considered to be local structures because they are associated with the point, $x.$

            In this work the fiber contents will be expanded as needed.  The fiber can include scaled structures for different types of numbers, different vector spaces, such as Hilbert spaces and products of these spaces. It can also include a representation of $M.$  A representation of $M$ at each point $x$ of $M$ is   defined  by a coordinate chart $\phi_{x}$ where \begin{equation}\label{phixM}\phi_{x}(M)=\mathbb{R}_{x}^{n}. \end{equation} Here $\mathbb{R}_{x}^{n}$ is a coordinate representation of $M$. Also $n=3$ (Euclidean) or $n=4$ (space time).

            The charts in the family, $\{\phi_{x}:x\epsilon M\},$ are consistent in that for  any two points $x,y$ and any point $z$ in $M,$ there are  tuples, $r_{x}^{n}$ and $r_{y}^{n}$, of real number values in $\mathbb{R}_{x}^{n}$ and in $\mathbb{R}_{y}^{n}$ such that \begin{equation}\label{phm1xy}
            \phi_{x}^{-1}(r_{x}^{n})=z=\phi_{y}^{-1}(r_{y}^{n}).\end{equation} In addition, $r_{x}^{n}$ must be a tuple of the same numbers  in $\mathbb{R}_{x}^{n}$ as $r_{y}^{n}$ is in $\mathbb{R}_{y}^{n}.$

            The fiber bundle with the contents described so far is given by \begin{equation}\label{MFSVR}
            \mathfrak{CVR}^{\cup}=M\times\bigcup_{s}(\bar{C}^{s} \times\bar{V}^{s})\times\mathbb{R}^{n}, \pi,M.\end{equation} The fiber at $x$ is given by \begin{equation}\label{pimm1x}\pi^{-1}(x)=\bigcup_{s}(\bar{C}^{s}_{x} \times\bar{V}^{s}_{x})\times\mathbb{R}^{n}_{x}.\end{equation} The bundle, $\mathfrak{CVR}^{\cup},$ is a principal fiber bundle in that the structures, $\bar{C}^{s} \times\bar{V}^{s}$ at the different levels, $s,$ are equivalent.  No value of $s$ is preferable over another.  This supported by the presence of a structure group $W_{CV}$ whose elements act freely and transitively on the structures at any fiber level, $s$, \cite{Drechsler}.   This is seen from the action of any element $W_{CV,d}$  of the structure group where\begin{equation}\label{WSVd} W_{CV,d}\bar{C}^{s} \times\bar{V}^{s}= \bar{C}^{ds} \times\bar{V}^{ds}.\end{equation} Note that $W_{CV,d}$ does not act on $\mathbb{R}^{n}$ as it is not scaled.

            Space and time dependent scaling of scalar and vector structures is accounted for by introduction of a smooth scalar scaling field, $g,$ on $M$. For each $x$ in $M,$ $g(x)$ is a complex number value in $\bar{C}.$ The field $g$ defines paths, $Q_{g},$ on the fiber bundle, with one path for each value of $s$ The paths are structure valued in that, for each $x$ in $M$, $Q_{g}(x)=\bar{C}^{sg(x)}_{x}\times\bar{V}^{sg(x)}_{x}.$ If $s=1$ which is the main case of interest, \begin{equation}\label{Qgx}Q_{g}(x)=\bar{C}^{g(x)}_{x}\times\bar{V}^{g(x)}_{x}. \end{equation}

             Scalar or vector  fields, $\psi_{sg}$, as sections on the fiber bundle, are defined by the requirement that for each $x$ in $M$ $\psi_{sg}(x)$ is a scalar in $\bar{C}^{sg(x)}_{x}$  or a vector in $\bar{V}^{sg(x)}_{x}$.  Since the results are independent of $s$, $s=1$ will be assumed in the following. The special case of no scaling has  $g(x)=1$ everywhere. Then  the field,  $\psi,$ is a level section on the bundle.

            \section{Connections}\label{C}
            The fact that for each $x$, $\psi_{g}(x)$ is a scalar or vector in the fiber at $x$, creates a  problem in that one cannot directly define integrals or derivatives of these fields over $M$. The reason is that these operations require mathematical addition or subtraction of field components in different fibers.  These operations are not defined between fibers.  They are defined only for structures at the same scaling level within a fiber.

            This is remedied by the use of connections to parallel transport field components between fibers at different locations. In gauge theory the connections, as elements of a unitary gauge group, express the freedom to choose bases for the vector spaces at each point of $M$ \cite{Yang,Montvay}.  Here the connections, as elements of the group, $GL(1,C)$, express the freedom to choose scaling factors for complex number and vector space structures at each point of $M$.

            The purpose of a connection is to  transport a field value $\psi_{g}(y),$ as a scalar or vector in $\bar{C}^{g(y)}_{y}$ or $\bar{V}^{g(y)}_{y}$  in the fiber at $y,$ to a value that can be combined with the value of $\psi_{g}(x)$ in the fiber at $x.$ This is clearly needed  because integrals or derivatives on $M$  require combinations of field values in different fibers.

            The connection that enables the combination of $\psi_{g}(y)$ with $\psi_{g}(x)$ has both horizontal and vertical components.  The horizontal component maps $\psi_{g}(y)$ to the same scalar or vector in $\bar{C}^{g(y)}_{x}$ or $\vec{V}^{g(y)}_{x}$ in the fiber at $x$.  The vertical component maps this vector to a scalar or vector  in $\bar{C}^{g(y)}_{g(x),x}$ or in $\bar{V}^{g(y)}_{g(x),x}.$  Here these scaled structures are given by Eqs. \ref{BCts} and \ref{BVts}.

            This can be expressed explicitly by defining the connection map $C_{g}(x,y)$ by
            \begin{equation}\label{Cgxy} C_{g}(x,y)\psi_{g}(y)\rightarrow \psi_{g}(y)^{y}_{x}\rightarrow \frac{g(y)_{g(x)}}{g(x)_{g(x)}}\psi_{g}(y)_{x}.\end{equation} Here $\psi_{g}(y)^{y}_{x}$  is the same scalar or vector   in  $\bar{C}^{g(y)}_{x}$ or $\bar{V}^{g(y)}_{x}$   as $\psi_{g}(y)$ is in $\bar{C}^{g(y)}_{y}$ or $\bar{V}^{g(y)}_{y},$ and $\psi_{g}(y)_{x}$  is the  same scalar or vector in $\bar{C}^{g(x)}_{x}$ or $\bar{V}^{g(x)}_{x}$ as $\psi_{g}(y)$ is in $\bar{C}^{g(y)}_{y}$ or $\bar{V}^{g(y)}_{y}.$ Also $g(y)_{g(x)}$ and $g(x)_{g(x)}$ are numbers in $\bar{C}^{g(x)}_{x}$ with values $g(y)$ and $g(x).$

            In the following, to avoid clutter, the subscripts, $g(x)$ will be suppressed from expressions with the scaling field, $g.$  If needed by the context, they will be inserted.

            If $y=x+d^{\mu}x$, as in a derivative, then \begin{equation}\label{fracgygx}\frac{g(y)}{g(x)} =\frac{g(x+d^{\mu}x)}{g(x)} =1+\frac{\partial_{\mu,x}g(x)}{g(x)}d^{\mu}x.\end{equation} A Taylor expansion has been used here on $g(x+d^{\mu}x).$

            It must be emphasized that this is a simplified description of the effect of the connection that skips over some details.  The connection can also be defined  as the product of  a pair of operators on the scalar  and vector structures.  Define the  structure operator, $W_{g}(x,y)$ by \begin{equation}\label{Wgxy}W_{g}(x,y)\bar{C}^{ds}_{y}\times\bar{V}^{ds}_{y}= \bar{C}^{ds}_{x}\times\bar{V}^{ds}_{x}.\end{equation} This operator is an isomorphism that  transports structures  at level $ds$ in the fiber at $y$ to  the  structures at the same level, $ds$, in the fiber at $x.$  The  components in in the transported structures are the same as they are in the original structures.  Numbers or vectors in the sets, $C_{y}$ or $V_{y},$ are mapped to the same numbers or vectors in $C_{x}$ or $V_{x}.$ Here $s=g(x)$ and $t=ds=g(y).$

            The operator $Z^{ds}_{s}$ of Eq. \ref{Zts} is used to map $\bar{C}^{ds}_{x}\times \bar{V}^{ds}_{x}$ to $\bar{C}^{ds}_{s,x}\times\bar{V}^{ds}_{s,x}.$  The definition of the connection, $C_{g}(x,y),$ as\begin{equation}\label{CZW}C_{g}(x,y)=Z^{ds}_{s}W_{g}(x,y) \end{equation}gives, \begin{equation}\label{CgxybC}C_{g}(x,y)(\bar{C}^{ds}_{y} \times\bar{V}^{ds}_{y})= \bar{C}^{ds}_{s,x} \times\bar{V}^{ds}_{s,x}.\end{equation} The components of $\bar{C}^{ds}_{s,x}$ and $\bar{V}^{ds}_{s,x}$ are given by Eqs. \ref{Zts} and \ref{BVts}.  Eq. \ref{Cgxy} follows directly from these results.

            It will be useful to represent $g$ as the exponential of a pair of real scalar fields $\alpha(x), \beta(x)$ as in \begin{equation}\label{gxe}g(x)=e^{\alpha(x)+i\beta(x)}=e^{\gamma(x)}.\end{equation}  In this case, \begin{equation}\label{fgyx}\frac{g(y)}{g(x)}=e^{\gamma(y)-\gamma(x)}.\end{equation}

            If $y=x+d^{\mu}x$ then \begin{equation}\label{egax}e^{\gamma(x+d^{\mu}x)-\gamma(x)} =e^{\partial_{\mu,x} \gamma(x)d^{\mu}x}=1+\vec{\Gamma}_{\mu}(x)d^{\mu}x.\end{equation}This result is obtained by use of a Taylor expansion of $\gamma(x+d^{\mu}x)$ followed by an expansion of the exponential to first order. The vector field, $\vec{\Gamma}$ is the gradient of $\gamma.$ Here $\vec{\Gamma}$  is a complex vector field as in
            \begin{equation}\label{vGvAvB}\vec{\Gamma}=\vec{A}+i\vec{B}\end{equation}where $\vec{A}$ and $\vec{B}$ are  gradients of the scalar $\alpha$ and $\beta$ fields.

            \section{Quantum Mechanics}\label{QM}
            Quantum mechanics provides many instructive examples to see the use of   the scaling field, $g$  in the description of localized quantum entities.  These include single and multiparticle wave functions as integrals over space and  momentum, kinetic energy operators as space derivatives, and two body interaction potentials.

            As noted earlier, justification for the localization of quantum entities is based on the observation that the value of a number at one point of space time does not determine the value of the same number at another point. As noted in Section \ref{FB} this leads to separate scaled scalar field structures at each space time point.  The structures are distinguished by scaling factors.  For complex number structures, the scaling factors are complex.

            Much of the earlier work, such as that in \cite{BenINTECH,BenNOVA}, applies this space time dependence of values of numbers  to other areas of physics.  This includes all quantities whose description includes rational, real, or complex numbers. It necessarily requires the localization of the description of physical quantities since the space time dependence of number values requires the localization of the number structures.

            Here this localization is applied to quantum mechanical quantities. The manifold, $M,$ will be taken to be  three dimensional Euclidean space for nonrelativistic quantum mechanics.

            \subsection{Single particle quantum mechanics}\label{SPQM}
            \subsubsection{Position Space representation}\label{PSR}
            As is well known, a single particle wave packet, $\psi$ can be represented by an integral as in
            \begin{equation}\label{wvepckt}\psi=\int\psi(y)|y\rangle dy.\end{equation} Here $\psi$ is a vector in a Hilbert space, $\bar{H},$ with $\psi(y)$ a complex number in $\bar{C}$.  The integral is over all of $M$.   In this representation, $\bar{H}$ and $\bar{C},$ as mathematical structures, and $\psi$  are not associated with any space locations or regions.

            As was noted, the space dependence of the values of complex numbers requires that this description be localized to one based on the local mathematical structures. The fiber bundle framework already discussed is useful for this purpose.

            A suitable fiber bundle for  these states is,\begin{equation} \label{MFCHR}\mathfrak{CHR}^{\cup}=M\times\bigcup_{c}(\bar{C}^{c} \times\bar{H}^{c}) \times\mathbb{R}^{3},\pi,M.\end{equation} The fiber at each point $x$ of $M$ is
            \begin{equation}\label{pimx}\pi^{-1}(x)=\bigcup_{c}(\bar{C}^{c}_{x} \times\bar{H}^{c}_{x}) \times\mathbb{R}^{3}_{x}.\end{equation}For each complex scaling value, $c,$ $\bar{H}^{c}$ is a Hilbert space suitable for expressing wave packet states as integrals over $\mathbb{R}^{3}.$ Here $\mathbb{R}^{3}=\phi(M)$ is a chart representation of $M$. The lifting of the wave packet state, $\psi$ of Eq. \ref{wvepckt} to an integral over $\mathbb{R}^{3}$ as a state in $\bar{H}^{c}$ is an example. In this case, the lifted amplitude, $\psi_{c}(z),$ is a complex number in $\bar{C}^{c}.$

            In this bundle the lifted states $\psi_{c}$ in the fiber are not associated with any point of $M$. The fibers at each point, $x$, of $M$ provide a simple localization in that $\psi_{c,x}$ is a local description of the wave packet in the fiber, $\pi^{-1}(x).$

            This description of wave packet localization does not take account of scalar scaling field in any nontrivial way.  This can be changed by treating wave packet states as integrals over a vector field that is a section on the bundle $\mathfrak{CHR}^{\cup}.$  This is much more in tune with the procedure followed in gauge theories.

            To achieve this one notes that the integrand of Eq. \ref{wvepckt} is a vector valued field over $M$.  This can be made more explicit by  expressing the wave packet integral by \begin{equation}\label{pslam}\psi=\int\lambda(y)dy.\end{equation} Here $\lambda$ is a vector valued field with values, $\lambda(y)=\psi(y)|y\rangle,$ in $\bar{H}$ for each $y$ in $M$.

            One now follows the prescription from gauge theory by treating $\lambda$ as a section on the fiber bundle. In the absence of scaling, for each $y,$ $\lambda(y)$ is a vector in the Hilbert space, $\bar{H}_{y},$ in the fiber at point $y$ in $M$. In the presence of the scaling field $g$, $\lambda(y)$ is a vector in $\bar{H}^{g(y)}_{y}$  at level $g(y)$ in the fiber at $y.$  This is the case of interest here.

            The properties of $\lambda$ as a section on the fiber bundle have the consequence that the integral of Eq. \ref{pslam} is not defined.  The reason is that, for each point $y$ in $M$, the integrand, $\lambda(y),$  is in $\bar{H}^{g(y)}_{y}$ in the fiber at $y$. Addition of the integrand is defined only for vectors in the same vector space in a fiber. It is not defined for vectors or scalars at different levels in different fibers.

            This is remedied by choice of a reference point, $x$, and use of the connection, defined by Eq. \ref{Cgxy}, to map integrands at points $y$ to integrands in the fiber at $x.$
            The effect is to replace the integral over $M$ by an equivalent local integral over $\phi_{x}(M)=\mathbb{R}^{3}_{x}.$  Each point $y$ in $M$ becomes a point $z_{y}$ in $\mathbb{R}^{3}_{x}$  where $z_{y}=\phi_{x}(y).$

            The connection first maps $\lambda(y)$ in $\bar{H}^{g(y)}_{y}$ to the  vector, $\lambda(y)_{x}$ in $\bar{H}^{g(y)}_{x}.$ Here $\lambda(y)_{x}$ is the same vector in $\bar{H}^{g(y)}_{x}$ as $\lambda(y)$ is in $\bar{H}^{g(y)}_{y}.$ This is followed by a level change map of $\lambda(y)_{x}$ to a scaled vector,  in $\bar{H}^{g(x)}_{x}.$  The net effect of the connection is described using Eq. \ref{CZW}. One has \begin{equation}\label{Cglamb}C_{g}(x,y)\lambda(y)= Z^{ds}_{s}W_{g}(x,y) \lambda(y)=Z^{ds}_{s}\lambda(y)_{x}= \frac{g(y)_{s}}{g(x)_{s}}\lambda(y)_{x}.\end{equation} Here $s=g(x)$ and $ds=g(y).$

            The description of the wave packet locally at $x,$ as an integral of a scaled vector field over $\mathbb{R}^{3}_{x},$  is facilitated by lifting the $g$ field into the fiber as a scaling field over $\mathbb{R}^{3}_{x}.$ The corresponding field, $g_{x},$ is defined by the requirement that for each $y$ in $M$, \begin{equation}\label{gxph}g_{x}(\phi_{x}(y))=g_{x}(z_{y})=g(y)_{s}. \end{equation}This equation says that $g_{x}(z_{y})$ is a number in $\bar{C}^{g(x)}_{x}$ that has value $g(y).$

            Eq. \ref{Cglamb} becomes \begin{equation}\label{Cglambx} C_{g}(z_{y},u_{y}) \lambda_{y}(u_{y})= \frac{g_{x}(z_{y})}{g_{x}(z_{x})}\lambda_{x}(z_{y}). \end{equation}The implied division operation in the ratio is that in $\bar{C}^{g(x)}_{x}.$  $\lambda_{y}$ and $\lambda_{x}$ denote the lifting of $\lambda$ to  vector fields over $\mathbb{R}^{3}_{y}$ and $\mathbb{R}^{3}_{x}.$ The point in $\mathbb{R}^{3}_{y}$ corresponding to $y$ in $M$ is $u_{y}$. Also  $\lambda_{x}(z_{y})$ is the same vector in $\bar{H}^{g(x)}_{x}$ as $\lambda_{y}(u_{y})$ is in $\bar{H}^{g(y)}_{y}$

\begin{figure}[h]\begin{center}\vspace{2cm}
            \rotatebox{270}{\resizebox{130pt}{130pt}{\includegraphics[170pt,260pt]
            [520pt,610pt]{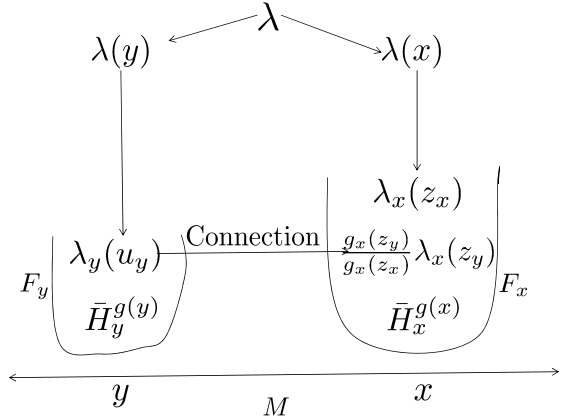}}}\end{center}\caption{Illustration of the steps in mapping of a vector field over $M$ to a local vector field over $\mathbb{R}^{3}_{x}$ in a fiber at $x.$ $\lambda(x)$ and $\lambda(y)$ are  values of the  vector field, at $x$ and $y.$ $\lambda_{y}(u_{y})$ and $\lambda_{x}(z_{x})$ are the corresponding  values in the fibers, $F_{y}$ and $F_{x}$ at $y$ and $x$. $u_{y}$ and $z_{x}$ are points in $\mathbb{R}^{3}_{y}$ and $\mathbb{R}^{3}_{x}$ that correspond  to $y$ and $x$ in $M$. The two connection steps that map $\lambda_{y}(u_{y})$ in $\bar{H}^{g(y)}_{y}$ to a vector in $\bar{H}^{g(y)}_{g(x),x}$ are shown by the horizontal arrow from fiber $F_{y}$ to $F_{x}.$ The effect of the connection is to multiply the the vector field value by $g_{x}(z_{y})/g_{x}(z_{x}).$ Here $g_{x}$ is the scaling field $g$ lifted to a local field, $g_{x}$ over $\mathbb{R}^{3}_{x}.$ and $\lambda_{x}(z_{y})$ is the same vector in  $\bar{H}^{g(x)}_{x}$ as $\lambda_{y}(u_{y})$ is in $\bar{H}^{g(y)}_{y}.$ Also $\phi_{x}(y)=z_{y},\;\phi_{x}(x)=z_{x},$ and $\phi_{y}(y)=u_{y}.$} \label{NSQM2}\end{figure}

            Figure \ref{NSQM2} shows schematically the steps described here for locations, $y$ and $x$ in $M$. Both the projection of $\lambda$ as a section on the fiber bundle, and the use of connections to transform section values in different fibers to a common fiber, are shown.

            The local scaled wave  packet, $\psi_{g,x}$ can be defined as an integral over $\mathbb{R}^{3}_{x}.$ From Eq. \ref{Cglambx} one has \begin{equation}\label{psig}\begin{array}{l}\psi_{g,x}= g_{x}(z_{x})^{-1} \int_{x}g_{x}(z)\lambda_{x}(z)dz=g_{x}(z_{x})^{-1}\int_{x} g_{x}(z)\psi_{x}(z) |z\rangle_{x} dz\\\\\hspace{1cm}=e^{-\gamma_{x}(z_{x})}\int_{x}e^{\gamma_{x}(z)} \psi_{x}(z)|z\rangle_{x}dz\end{array}\end{equation} as the final local expression for the wave packet with scaling included.  Here Eq. \ref{fgyx} has been used.

            Eq. \ref{psig} shows explicitly that $\psi_{g,x}$ is the local expression, in the fiber at $x$, of the original global wave packet of Eq. \ref{wvepckt}. The localization includes scaling. This is shown by the exponential factors, $e^{\gamma_{x}(z)-\gamma_{x}(z_{x})}.$ Note that $\psi_{x}$ is the localization of $\psi$ with $g$  a constant field. This is equivalent to no scaling effect on the wave packet.

            The location, $x,$ of the reference fiber can be changed. The wave packet, $\psi_{g,w}$ for another fiber location, $w$ is obtained by replacing the subscript $x$ in Eq. \ref{psig} everywhere by $w.$  One can describe this by defining an operation on the fibers that maps the fiber contents at one location onto those at another location on $M.$   This in essence a translation operation on the fibers.

            Let $\mathcal{M}(w,x)$ be an operator that maps contents of the fiber at $x$ onto those at $w.$ The action of $\mathcal{M}(w,x)$ on the state, $\psi_{g,x}$ is defined by \begin{equation}\label{psigw} \begin{array}{l} \mathcal{M}(w,x)\psi_{g,x}= g_{w}(z_{w})^{-1}\int_{w} \mathcal{M}(w,x)(g_{x}(z) \psi_{x}(z) |z\rangle_{x})dz\\\\\hspace{1cm}=e^{\gamma_{w}(z_{w})}\int_{w}e^{-\gamma_{w}(z)} \psi_{w}(z)|z\rangle dz=\psi_{g,w}.\end{array} \end{equation} Here $z_{w}$ is the point in $\mathbb{R}^{3}_{w}$  defined by $\phi_{w}(w)=z_{w}.$ The integral is over all points in $\mathbb{R}^{3}_{w}.$ Also $\gamma_{w}$ and $\psi_{w}$ are the same $\bar{C}^{g(w)}_{w}$ valued fields over $\mathbb{R}^{3}_{w}$ as $\gamma_{x}$ and $\psi_{x}$ are as $\bar{C}^{g(x)}_{x}$ valued fields over $\mathbb{R}^{3}_{x}.$

            Note!!  $\mathcal{M}(w,x)$ describes a map or translation of the local mathematical description of the state, $\psi,$ at $x$ to that at $w.$  It does not correspond to a physical translation of the state  from $x$ to $w.$ The location of  $\psi_{w}$ on $\mathbb{R}^{3}_{w}$ is the same as the location of $\psi_{x}$ on $\mathbb{R}^{3}_{x}.$

            \subsubsection{Momentum space representation}\label{MSR}
            As is well known in quantum mechanics,  states  can be expressed as wave packets over either position space or momentum space. These representations are equally valid.  Also they can be transformed into one another.  Fiber bundles, along with connections, can be defined separately for each of the two types of representations. One representation has  fibers at momentum positions, $p,$ in a momentum space manifold.  The other representation has fibers at positions, $x,$ in a position space manifold. One can also describe position representations of a wave packet in the fiber at momentum $p$. Conversely one can describe  momentum representations of wave packets in a fiber at position, $x$.  An example of this is the momentum representation of $\psi_{g,x}$ in a bundle fiber at $x.$

            A suitable fiber bundle for this example is given by\begin{equation}\label{MFMCHPR} \mathfrak{CHPR}^{\cup}=M\times\bigcup_{c}(\bar{C}^{c}\times\bar{H}^{c})\times \mathbb{P}^{3} \times \mathbb{R}^{3},\pi,M.\end{equation}Here $M$ is the position space manifold. The fiber at $x$ is given by \begin{equation}\label{pim1xp}\pi^{-1}(x)=\bigcup_{c}(\bar{C}^{c}_{x} \times\bar{H}^{c}_{x})\times\mathbb{P}^{3}_{x}\times\mathbb{R}^{3}_{x}.\end{equation}Here $\mathbb{P}^{3}_{x}$ is the  representation of momentum space in the fiber at $x.$

            The position space representation of $\psi_{g,x}$ in the fiber at $x$ is given by Eq. \ref{psig}.  As before $g$ is the  scaling field. The momentum representation of $\psi_{g,x}$ is obtained by use of momentum completeness relations in Eq. \ref{psig} to obtain \begin{equation}\label{psigxpp}\psi_{g,x}=\frac{e^{-\gamma_{x}(z_{x})}}{(2\pi\bar h)^{3}}\int_{x}\int_{x}dp_{x}dq_{x}|p_{x}\rangle\langle p_{x}| e^{\gamma_{x}(z)}|q_{x}\rangle\langle q_{x}|\psi_{x}\rangle.\end{equation} The contribution of $e^{\gamma_{x}}$ to the momentum distribution is given by \begin{equation}\label{ppgp}\langle p_{x}|e^{\gamma_{x}}|q_{x}\rangle = \int_{x}\langle p_{x}|z\rangle e^{\gamma_{x}(z)}\langle z|q_{x}\rangle dz = \int_{x}e^{\frac{\mbox{\small $i(p_{x}-q_{x})z$}} {\mbox{\small $\hbar$}}}e^{\gamma_{x}(z)} dz.\end{equation} The subscript,  $x$ on the momentum variables indicate that they are momenta in $\mathbb{P}^{3}_{x}$ in the fiber at $x.$ The amplitude for the unscaled state in the fiber at $x$ to have momentum $q_{x}$ is $\psi_{x}(q_{x}).$

            This result shows that, for the momentum representation of $\psi_{g,x}$ in the fiber at $x$, one cannot define a momentum representation of the scaling field, $g_{x},$ that is independent of $\psi.$ The two are connected by the convolution integral over $q_{x}$. This is quite different from the space representation where the scaling factor and $\psi$ appear as a product of $e^{\gamma_{x}(z)}$ and $\psi_{x}(z)$.

            If both the real and imaginary parts of $\gamma_{x}(z)$ are very small for relevant values of $z,$ then one can write $e^{\gamma_{x}(z)}\approx 1+\gamma_{x}(z).$ This gives \begin{equation} \label{intxefrac}\int_{x}e^{\frac{\mbox{\scriptsize $i(p-q)z$}}{\mbox{\scriptsize $\hbar$}}}e^{\gamma_{x}(z)} dz=\int_{x}e^{\frac{\mbox{\scriptsize $i(p-q)z$}} {\mbox{\scriptsize $\hbar$}}}(1+\gamma_{x}(z)) dz =(2\pi\hbar)^{3}\delta(p-q)+\int_{x}e^{\frac{\mbox{\scriptsize $i(p-q)z$}}{\mbox{\scriptsize $\hbar$}}}\gamma_{x}(z) dz.
            \end{equation}

            \subsubsection{Local nonrelativistic quantum mechanics}

           It is evident that nonrelativistic quantum mechanics can be described locally in the fibers at different positions in $M$.  Additional mathematical structures can be added to the fibers as needed.

           In absence of scaling, momentum  and kinetic energy operators have their usual definitions  inside a fiber. In the fiber at $x$  they are defined by \begin{equation}\label{tilpx}\tilde{p}_{x}= -i\hbar\sum_{j=1}^{3} \frac{\partial}{\partial z^{j}}\mbox{ and }\tilde{K}_{x}(z)= \frac{\tilde{p_{x}}^{2}}{2m}= -\frac{{\hbar}^{2}}{2m} \sum_{j=1}^{3} \frac{\partial^{2}}{(\partial z^{j})^{2}}.\end{equation}The derivatives in these operators are defined on $\mathbb{R}^{3}_{x}.$

            The operators in the fiber at $x$ can be combined with potentials  in the fiber at $x$ to define Hamiltonians.   For one body potentials the Hamiltonian  has the usual form \begin{equation} \label{tilHx}\tilde{H}_{x}(z)=\tilde{K}_{x} (z)+\tilde{V}_{x}(z).\end{equation} The Schr\"{o}dinger equation has the usual form \begin{equation}\label{frhbi}i\hbar\frac{d}{dt}\psi_{x}=\tilde{H}_{x} \psi_{x}.\end{equation}

            In the absence of scaling, Eqs. \ref{tilpx} are valid even though the momentum operator, as a space derivative, is the local limit of a nonlocal operation on $M$.   This also describes the special case in which the scaling field, $g$ is constant everywhere. For more general cases where $g$ is not constant, the $g$ field does affect the momentum.  This can be seen from the usual definition of the action of the momentum operator on a wave packet as\begin{equation}\label{tilp}\tilde{p}\psi= -i\hbar\sum_{j}\frac{\partial} {\partial x^{j}}\psi.\end{equation}This derivative is defined on $M$.

            Mapping this definition into the local mathematical structures in the fiber bundle $\mathfrak{CHPR}^{\cup}$ has the same problem as does the definition of derivatives in gauge theory Lagrangians. The  subtraction of $\psi_{g}(x+dx^{j})$ from $\psi_{g}(x),$ implied in the definition of the derivative \begin{equation}\label{partial}\frac{\partial}{\partial x^{j}}\psi_{g}= \frac{\psi_{g}(x+dx^{j}) -\psi_{g}(x)}{dx^{j}},\end{equation}(limit $dx^{j}\rightarrow 0$ implied) is not defined because the terms to be subtracted are in different fibers. The vector $\psi_{g}(x+dx^{j})$ is in the Hilbert space, $\bar{H}^{g(x+dx^{j})}_{x+dx^{j}}$ and $\psi_{g}(x)$ is a vector in $\bar{H}^{g(x)}_{x}.$

            This is remedied by use of the $g$ field connection, defined in Eq. \ref{Cglamb}, to parallel transform $\psi_{g}(x+dx^{j})$ to the fiber at $x$.  The resulting derivative is given by \begin{equation} \label{Djx}D_{j,x}\psi_{g}=\frac{\frac{\mbox{\small $g(x+dx^{j})$}}{\mbox{\small $g(x)$}}\psi_{g}(x+dx^{j})_{x}-\psi_{g}(x)}{dx^{j}}. \end{equation} Here $\psi_{g}(x+dx^{j})_{x}$ is the same vector in $\bar{H}^{g(x)}_{x}$ as $\psi_{g}(x+dx^{j})$ is in $\bar{H}^{g(x+dx^{j})}_{x+dx^{j}}.$

            The  Taylor expansion of $g(x+dx^{j})$ gives \begin{equation} \label{Djxps} D_{j,x}\psi=\partial_{j,x} \psi+\frac{\partial_{j,x}g(x)}{g(x)}\psi.\end{equation}
            Use of Eqs.  \ref{fgyx} and \ref{egax} gives \begin{equation} \label{Djxpsi} D_{j,x}\psi_{g}=\partial_{j,x} \psi_{g}+\Gamma_{j}(x)\psi_{g}\end{equation} where \begin{equation}\label{Gam}\partial_{j,x}\gamma=\Gamma_{j}(x).\end{equation}

            This result shows that localization of the global momentum operator $\tilde{p}$ to a local operator, $\tilde{p}^{x},$ in the fiber  at a point $x$ of $M$ introduces a gradient vector field according to \begin{equation}\label{tildp}\tilde{p}\rightarrow \tilde{p}^{x}=\tilde{p}_{x}-i\hbar \vec{\Gamma}_{x}=\tilde{p}_{x}-i\hbar \vec{A}_{x}+\hbar\vec{B}_{x}.\end{equation} Here $\vec{\Gamma}_{x}$ is the field in the fiber at $x$ that corresponds to the lifting of $\vec{\Gamma}$ as a vector field on $M$ to a vector field on $\phi_{x}(M)=\mathbb{R}^{3}_{x}.$  It is also the gradient field of $\gamma_{x}.$ Also $\tilde{p}_{x}$ is the usual expression for the momentum in the fiber at $x$. Eq. \ref{tildp} is the same expression one obtains in quantum mechanics where the momentum in the presence of an external field is replaced by a canonical momentum.\footnote{\label{f1}The same result as in Eq. \ref{tildp} can be obtained by taking the derivative of $e^{\gamma_{x}(z)}\psi(z)$ in the fiber at $x.$}

            The kinetic energy operator is treated in the same way as is the momentum operator. The usual global representation, $\tilde{K}=\vec{p}\cdot\vec{p}/2m,$ has the local expression in the fiber at $x$ as \begin{equation}\label{Kx}\tilde{K}^{x}=\frac{(\tilde{p}^{x})^{2}}{2m}=\frac{(\tilde{p}_{x}-i\hbar \vec{\Gamma}_{x})^{2}}{2m} =\tilde{K}_{x}+ \frac{-i\hbar\tilde{p}_{x}\vec{\Gamma}_{x}-i\hbar\vec{\Gamma}_{x} \tilde{p}_{x}- \hbar^{2}\vec{\Gamma}_{x}^{2}}{2m}.\end{equation}

            The usual component expansion gives \begin{equation} \label{Ksupx}\tilde{K}^{x}(z)=(\frac{-\hbar^{2}}{2m})_{x}\sum_{j=1}^{3}[\partial^{2}_{j,z}+
            \partial_{j,z}(\Gamma_{x,j}(z))+2\Gamma_{j,x}(z)\partial_{j,z}+\Gamma_{x,j}(z)^{2}]. \end{equation} This result makes use of the commutation relation\begin{equation}\label{comm}p_{j} \Gamma_{x,j}-\Gamma_{x,j}p_{j}=-i\hbar\frac{d\Gamma_{x,j}}{d^{j}z}.\end{equation} Here \begin{equation} \label{frdGj}\frac{d\Gamma_{x,j}}{d^{j}z}=\frac{d^{2}\gamma_{x}}{(d^{j}z)^{2}}. \end{equation}

            The usual global representation of the action of a Hamiltonian on a one particle state is given by $\tilde{H}\psi.$  The space representation of this as an integral over $M$ is given by
            \begin{equation}\label{Hpsi}\tilde{H}\psi=\int\tilde{H}(y)\psi(y)dy=\int(\tilde{K}(y)+\tilde{V}(y))
            \psi(y)dy.\end{equation}In the presence of scaling, localization of this quantity to a fiber at $x$ gives an expression similar to that for $\psi_{g,x}$ in Eq. \ref{psig}.  One has \begin{equation} \label{Hpsix}(\tilde{H}\psi)_{g,x}=e^{-\gamma_{x}(z_{x})}\int_{x}e^{\gamma_{x}(z)}\tilde{H}^{x}(z)\psi_{x}(z)dz.
            \end{equation}The kinetic energy operator in $\tilde{H}^{x}(z)$ is given by Eq. \ref{Ksupx}. For each $z$ in $\mathbb{R}^{3}_{x},$ $\psi_{x}(z)$ is a vector in $\bar{H}^{g(x)}_{x}.$

            This equation expresses the localization with scaling of the global expression $\tilde{H}\psi.$  It shows the action of an operator on $\psi$ before scaling. Another way to approach localization of global expressions is to commute localization with the operator action and consider the action of the local Hamiltonian in the fiber at $x$ on the localized, global wave function.  This is represented by the expression \begin{equation}\label{tHxp}\tilde{H}_{x}\psi_{g,x}=\tilde{H}_{x} e^{\gamma_{x}}\psi_{x}=e^{-\gamma_{x}(z{x})}\int_{x}\tilde{H}_{x}(z) e^{\gamma_{x}(z)}\psi_{x}(z)dz \end{equation}In this equation $\tilde{H}_{x}(z)$ has the usual form given in Eq. \ref{tilHx}.

            Evaluation of the action of the kinetic energy term of $\tilde{H}_{x}$ on $e^{\gamma_{x}(z)} \psi_{x}(z)$ gives the result that\begin{equation}\label{tHxe}\tilde{H}_{x}e^{\gamma_{x}(z)} \psi_{x}(z)=e^{\gamma_{x}(z)}\tilde{H}^{x}(z)\psi_{x}(z).\end{equation}This shows that Eqs. \ref{Hpsix} and \ref{tHxp} are equivalent in that \begin{equation}\label{tHp}(\tilde{H}\psi)_{g,x} =\tilde{H}_{x}\psi_{g,x}.\end{equation}

            It follows that, even in the presence of scaling, localization of the global representation of $\tilde{H}\psi$ gives the same result as does the action of the local representation of the Hamiltonian, $\tilde{H}_{x}$ on the localized, global state, $\psi_{g,x}.$ In other words, the action of the Hamiltonian on a state vector commutes with localization. If $L_{g}(x)$ represents the localization operation to a point, $x$, of $M$, then these results show that \begin{equation} \label{LgxtH}L_{g}(x)\tilde{H}\psi=\tilde{H}_{x}L_{g}(x)\psi.\end{equation}

            \subsection{Multiparticle quantum mechanics}\label{mqm}

             The treatment of multiparticle quantum mechanics requires an expansion of the fiber bundle framework used here as well as introducing new features. This is especially so for entangled states and states of interacting systems.  The simplest case to consider is that of the quantum state for two interacting particles.  These can be either bosons, or fermions in different  states.

            \subsubsection{Two particle entangled states}\label{TPES}

            The effect of scaling on states describing the interaction of two  particles is different from the results for two noninteracting systems.  To see this let $\psi_{1,2}$ be an entangled state of two systems.  The usual  expression for this state is given by \begin{equation}\label{psi12ent}\psi_{1,2}=\int \psi_{1,2}(x,y)|x\rangle |y\rangle dxdy. \end{equation} Here $\psi_{1,2}(x,y)$ is the amplitude for finding  a particle at each of the two sites, $x$ and  $y.$ The integral is over pairs of points in $M$ and $\psi_{1,2}$ is a vector in the global Hilbert space $\bar{H}_{1,2}$ with associated scalars, $\bar{C}.$ Here $\bar{H}_{1,2}=\bar{H}_{1}\bigotimes\bar{H}_{2}$ is the two particle space that is the tensor product of $\bar{H}_{1}$ and $\bar{H}_{2}.$

            A simple example of an entangled two particle state is a Slater determinant state as
            \begin{equation}\label{Sd12}\psi_{1,2}(x,y)=\frac{1}{\sqrt{2}}(\psi_{1}(x)\psi_{2}(y)-\psi_{1} (y)\psi_{2}(x)).\end{equation} This antisymmetric state is suitable for fermions. For bosons the minus sign is replaced by a plus sign.

            In this state the entanglement is between spatial degrees of freedom.  States where there is entanglement  in spin degrees of freedom are exemplified by Bell states, such as
            \begin{equation}\label{psi12sp}\psi_{1,2,Sp}(x,y)=\frac{1}{\sqrt{2}}(|+\rangle_{1}|-\rangle_{2} -|-\rangle_{1}|+\rangle_{2})\psi_{1,2}(x,y).\end{equation}Here $\psi_{1,2}(x,y)$ can also be spatially entangled such as in a determinant or it can be a simple product state such as $\psi_{1}(x)\psi_{2}(y).$ Here spin entanglement will not be considered as the main concern is with spatially entangled states.

            As was done in Eq. \ref{pslam} for single particle states, $\psi_{1,2}$ can be written as
            \begin{equation}\label{pl12}\psi_{1,2}=\int\lambda_{1,2}(x,y)dxdy.\end{equation}Here $\lambda_{1,2}$ is a vector valued field from pairs of points in $M$ to vectors in $\bar{H}_{1,2}.$

            The goal is to define a fiber bundle so that $\lambda_{1,2}$ becomes a section on the bundle.  In this case, for each point pair, $x,y$, $\lambda_{1,2}(x,y)$   is a vector in the  two particle Hilbert space in the fiber associated with the pair, $x,y$.

            This can be done by generalizing the fiber bundle description for single particle states to accommodate two particle states.  A suitable bundle is given by\begin{equation}\label{MCHR2} \mathfrak{CHPR}^{\cup}_{2} =M\times\bigcup_{c}(\bar{C}^{c} \times\bar{H}_{1,2}^{c}) \times\mathbb{P}^{3}\times\mathbb{R}^{3}, \pi,M.\end{equation}

            The projection operator, $\pi,$ is different from that in the previously described bundles in that it projects onto pairs of points in $M.$  Conversely the fibers are associated with pairs of points of $M$ instead of single points. The fiber associated with the point pair, $x,y$ is given by \begin{equation}\label{pim1xy}\pi^{-1}(x,y)= \bigcup_{c}(\bar{C}^{c}_{x,y}\times \bar{H}^{c}_{1,2,x,y})\times\mathbb{P}^{3}_{x,y}\times\mathbb{R}^{3}_{x,y}.\end{equation}
            Here $$\bar{H}^{c}_{1,2,x,y}=(\bar{H}_{1}\bigotimes\bar{H_{2}})^{c}_{x,y}$$ and  $$\mathbb{R}^{3}_{x,y}=\phi_{x,y}(M).$$ Also $\mathbb{P}^{3}_{x,y}$ is the conjugate momentum space, and $\phi_{x,y}$ is a chart in the family of consistent charts, one for each pair of points in $M.$ Chart consistency is defined by the requirement that for each two point pairs, $v,w$ and $x,y$ in $M$, $\phi_{v,w}(\phi_{x,y}^{-1}(z))$ is the same number tuple in $\mathbb{R}^{3}_{v,w}$ as $z$ is in $\mathbb{R}^{3}_{x,y}.$ This must hold for all $z$ in $\mathbb{R}^{3}_{x,y}.$

            The scaling field, $g_{2}(x,y)$ is a generalization of $g(x)$ to a scalar field over pairs of locations in $M.$ A reasonable choice for the values of $g_{2}(x,y)$ is  the geometric average of the values of $g(x)$ and $g(y).$  That is \begin{equation}\label{gxysq}g_{2}(x,y)=\sqrt{g(x)g(y)}=e^{\gamma_{2}(x,y)}. \end{equation} Here \begin{equation}\label{gxygx}\gamma_{2}(x,y)=\frac{\gamma(x)+\gamma(y)}{2} \end{equation} is the arithmetic average of $\gamma(x)$ and $\gamma(y).$. Note that $g_{2}(x,x)=g(x).$ From now on the subscript $2$ on $g$ and $\gamma$ will be suppressed unless required by the context.

            One now follows the procedure used for one particle states. The integrand, $\lambda_{1,2}$ is treated as a section on the fiber bundle, $\mathfrak{CHPR}^{\cup}_{2}$. In this case, the integral of $\lambda_{1,2}(x,y)$, as an integral over different fibers,  is undefined.

            This problem is fixed by choice of a reference pair, $v,w,$ of locations on $M$ and use of a connection to parallel transform the values of the integrand  in the fiber  at each point pair, $y,x,$ to the fiber at the reference point pair.   The effect of the connection is expressed as a generalization of that in Eq. \ref{Cglambx} to \begin{equation}\label{Cglamb12}C_{g}(v,w:x,y) \lambda(u_{x},u_{y})=\frac{g_{v,w}(z_{x},z_{y})}{g_{v,w}(z_{v},z_{w})}\lambda(z_{x},z_{y}).\end{equation} Here $$\lambda(z_{x},z_{y})=\psi_{1,2}(z_{x},z_{y})|z_{x},z_{y}\rangle.$$   The point pairs $u_{x},u_{y}$ in $\mathbb{R}^{3}_{x,y}$, and $z_{x},z_{y}$  and $z_{v},z_{w}$ in $\mathbb{R}^{3}_{v,w}$ are the respective $x,y$    chart values of $x,y,$ and the $v,w$ chart values of $x,y$ and $v,w.$ Also $u_{x},u_{y}$ is the same pair of tuples in $\mathbb{R}^{3}_{x,y}$ as $z_{x},z_{y}$ is in $\mathbb{R}^{3}_{v,w}.$

            In Eq. \ref{Cglamb12}  $g_{v,w}$ is the lifting of $g_{2}$ to the local, $\bar{C}^{g(v,w)}_{v,w}$ valued scaling field defined over $\mathbb{R}^{3}_{v,w}.$ This field can be defined as the geometric product of the lifting of the $g$ field to the fiber at $v,w$. One obtains \begin{equation} \label{g2vw}g_{v,w}(z_{y},z_{x})=\sqrt{g_{v,w}(z_{y})g_{v,w}(z_{x})}=e^{\gamma_{v,w}(z_{y},z_{x})}
            \end{equation}where \begin{equation}\label{gvwzy}\gamma_{v,w}(z_{y},z_{x})=\frac{\gamma_{v,w} (z_{y})+\gamma_{v,w}(z_{x})}{2}.\end{equation}

            These results can be used to give the definition of the scaled wave packet in the fiber at $v,w.$ The result is \begin{equation}\label{psig12}\begin{array}{l} \psi_{g,v,w}=g^{-1}_{v,w}(z_{v},z_{w}) \int_{v,w}g_{v,w} (z_{x},z_{y})\psi_{g,v,w}(z_{x}, z_{y})|z_{x},z_{y}\rangle_{1,2} dz_{x}dz_{y}\\\\\hspace{1cm} =e^{-\gamma_{v,w}(z_{v},z_{w})} \int_{v,w} e^{\gamma_{v,w}(z_{x},z_{y})} \psi_{v,w}(z_{x},z_{y})|z_{x},z_{y}\rangle_{1,2} dz_{x}dz_{y}.\end{array}\end{equation}Here, and for the rest of this subsection, the subscripts, $1,2$  on $\psi$ are suppressed.

            Restriction of the pair $v,w$ to $v,v$ achieves the result of mapping the global entangled state $\psi_{1,2}$ to a fiber at a single  reference location, $v$, of $M$. This follows from the fact that a fiber at $v,v$ in $M$ is equivalent to a fiber at $v.$ In this case Eq. \ref{psig12} simplifies to \begin{equation}\label{psivv}\psi_{g,v,v} =e^{-\gamma_{v}(z_{v})} \int_{v} e^{\gamma_{v,v}(z,z')} \psi_{v}(z,z')|z,z'\rangle dzdz'.\end{equation} Here \begin{equation} \label{gavvzz}\gamma_{v,v}(z,z')=\frac{1}{2}(\gamma_{v}(z)+\gamma_{v}(z')).\end{equation}The integral of Eq. \ref{psivv} is over pairs of points in $\mathbb{R}^{3}_{v,v}=\mathbb{R}^{3}_{v}.$

            \subsubsection{Momentum space representation}

            The momentum representation for the state, $\psi_{g,v,w}$ is an extension of that for the one particle state. From Eq. \ref{psig12} one has \begin{equation}\label{psig12vw} \psi_{g,v,w}= \frac{e^{\mbox{\small $-\gamma_{v,w}(z_{v},z_{w})$}}}{(2\pi\hbar)^{3}}\int_{v,w}dpdq|p,q\rangle\int_{v,w}\langle p,q| e^{\mbox{\small $\gamma_{v,w}(z,z')$}} \psi_{g,v,w}(z,z')|z,z'\rangle dzdz'.\end{equation}One can describe the momentum representation of the scaling factor and state by means of a convolution integral.

            This is done by expanding Eq. \ref{psig12vw} to \begin{equation}\label{psig12vwp}\begin{array}{l} \psi_{g,v,w}= \frac{\mbox{\small $e^{\mbox{\small $-\gamma_{v,w}(z_{v},z_{w})$}}$}}{\mbox{\small $(2\pi\hbar)^{6}$}}\int_{v,w}dpdq|p,q\rangle \int_{v,w}dzdz'\int_{v,w}dudu'\\\\\hspace{1cm}\times\int_{v,w}dp'dq'\langle p,q|u,u'\rangle e^{\gamma_{v,w}(u,u')}\langle u,u'|p',q'\rangle\langle p',q'| \psi_{1,2:v,w}(z,z')|z,z'\rangle. \end{array}\end{equation}Passing the $z,z'$ integral to the right allows the simplification of this expression to give \begin{equation}\label{psig12vwpq} \psi_{g,v,w}= e^{\mbox{\small $-\gamma_{v,w}(z_{v},z_{w})$}}\int_{v,w}dpdq|p,q\rangle\int_{v,w}dp'dq' g_{v,w}(p-p',q-q')
           \psi_{1,2:v,w}(p',q') .\end{equation}In this equation \begin{equation}\label{gpp'qq'}g_{v,w}(p-p',q-q')= \frac{1}{(2\pi\hbar)^{3}}\int_{v,w}dudu'e^{\frac{\mbox{\small $i(p-p')u$}}{\mbox{\small $\hbar$}}}e^{\frac{\mbox{\small $i(q-q')u'$}} {\mbox{\small $\hbar$}}}e^{\mbox{\small $\gamma_{v,w}(u,u')$}} \end{equation} and\begin{equation}\label{psppqp}\psi_{v,w}(p',q')=\frac{1}{(2\pi\hbar)^{3}}\int_{v,w}dzdz' e^{\frac{\mbox{\small $ip'z+iq'z'$}}{\mbox{\small $\hbar$}}}\psi_{v,w}(z,z').\end{equation}

           \subsubsection{local nonrelativistic quantum mechanics}
           Here some aspects of quantum mechanics in fibers at point pairs in $M$  are described.  The  fiber is is called local even though it is associated with two points of $M$ rather than just one.

           The action of the two particle momentum operator, $\tilde{p}_{1,2},$  at $x,y$ on a scaled entangled state component at $x,y$ is given by \begin{equation}\label{tp12}\begin{array}{l}\tilde{p}_{1,2}\psi_{1,2,g}(x,y)= (\tilde{p}_{1}+\tilde{p}_{2})\psi_{1,2,g}(x,y) =-i\hbar(\tilde{p}^{x}_{1}+\tilde{p}^{y}_{2})e^{\gamma(x,y)}\psi_{1,2}(x,y)\\\\\hspace{1cm}= e^{\gamma(x,y)}(\tilde{p}_{1,x}+\tilde{p}_{2,y}-i\hbar\frac{\mbox{\small $\vec{\Gamma}(x)+\vec{\Gamma}(y)$}}{\mbox{\small $2$}})\psi_{1,2}(x,y).\end{array} \end{equation}Here  $\tilde{p}_{1,x}$ and $\tilde{p}_{2,y}$ are taken respectively to be the usual derivative expressions  at $x$ and $y$.

           This expression corrects for the effects of the $g$ scaling field on the  two particle momentum operator.   It is not a description of the momentum inside a fiber.  This is obtained by lifting the expression for $\tilde{p}_{1,2}$ to the corresponding expression in the fiber at $v,w$  for the locations, $u,z$ in $\mathbb{R}^{3}_{v,w}.$ The result is given by \begin{equation}\label{tildpvw}\tilde{p}^{v,w}_{1,2}(u,z)=\tilde{p}^{v,w}_{1} (u)+ \tilde{p}^{v,w}_{2}(z) =\tilde{p}_{1,v,w}(u)+\tilde{p}_{2,v,w}(z)- i\hbar\frac{\vec{\Gamma}_{v,w}(u) +\vec{\Gamma}_{v,w}(z)}{2}.\end{equation}  Here \begin{equation}\label{tilpvw1}\tilde{p}^{v,w}_{1} (u)=\tilde{p}_{1,v,w}(u) - i\hbar\frac{\vec{\Gamma}_{v,w}(u)}{2},\;\mbox{ and }\;\tilde{p}^{v,w}_{2} (z)=\tilde{p}_{2,v,w}(z) - i\hbar\frac{\vec{\Gamma}_{v,w}(z)}{2}. \end{equation}In this equation $\Gamma_{v,w}(u)$ and $\Gamma_{v,w}(z)$ are the gradients of $\gamma_{v,w}$ at $u$ and $z.$

           The relation between the two particle momentum in the fiber at $v,w$ and that given for the points, $x,y$ in Eq. \ref{tp12} is given by the requirement that for each $x,y$ in $M,$ the momentum value of Eq. \ref{tp12} is the same as that of $\tilde{p}^{v,w}_{1,2}(u,z).$ The point pair, $u,z,$  is related to $x,y$ by $\phi_{v,w}(x)=u$ and $\phi_{v,w}(y)=z.$

           The global expression for the action of a two body Hamiltonian on a state, $\psi_{1,2}$ is given by
           \begin{equation}\label{H12p}\tilde{H}_{1,2}\psi_{1,2} =\int\tilde{H}(x,y)\psi(x,y)dxdy =\int(\tilde{K}_{1}(x)+\tilde{K}_{2}(y)+\tilde{V}(x,y))\psi_{1,2}(x,y)dxdy.\end{equation} The integral is over point pairs in $M$.

           Following the same procedure as was used for Eq. \ref{Hpsix},  use of connections to localize $\tilde{H}\psi$ to a fiber at $v,w$ with scaling included gives\begin{equation} \label{Hpsivw} (\tilde{H}\psi)_{vw}=e^{-\gamma_{v,w}(z_{v},z_{w})}\int_{v,w}e^{\gamma_{v,w}(u,z)}
           \tilde{H}^{v,w}(u,z)\psi_{v,w}(u,z)dudz.\end{equation}The kinetic energy operator for the two particles in $\tilde{H}^{v,w}(u,z)$ is given by \begin{equation} \label{tilKvw}\tilde{K}^{v,w}_{1}(u)+\tilde{K}^{v,w}_{2}(z)=\frac{\hbar^{2}}{m_{1}} (\tilde{p}^{v,w}_{1}(u))^{2}+\frac{\hbar^{2}}{m_{2}}(\tilde{p}^{v,w}_{2}(z))^{2}.\end{equation}The two body potential becomes $V_{v,w}(u,z)$ in the fiber.

           \subsubsection{Multiparticle entangled states}\label{MES}

           The  fiber bundle description for two particle entangled states in nonrelativistic quantum mechanics can be extended to entangled states of $n$ particles. A bundle for these states with scaling included is given by
           \begin{equation}\label{MCHPRn}\mathfrak{CHPR}_{n}^{\cup}=M\times\bigcup_{c}(\bar{C}^{c} \times\bar{H}_{n}^{c})\times\mathbb{P}^{3}\times\mathbb{R}^{3},\pi_{n},M.\end{equation} In this definition $\bar{H}^{c}_{n}$ is the $n$ fold tensor product of single particle scaled Hilbert spaces, all with the same scaling factor, $c,$ and $\pi_{n}$ is a one-one map of the fiber and $n$ tuples of points in $M$ to $n$ tuples of points in $M$. For each $n$ tuple, $x_{1},\cdots,x_{n}$, of points in $M$, \begin{equation}\label{pim1n}\pi^{-1}_{n}(x_{1,n})= \bigcup_{c}(\bar{C}^{c}_{x_{1,n}}\times \bar{H}^{c}_{x_{1,n}})\times\mathbb{P}^{3}_{x_{1,n}} \times\mathbb{R}^{3}_{x_{1,n}}.\end{equation}Here $x_{1,n}$ is short notation for $x_{1},x_{2},\cdots,x_{n}.$

           An $n$ particle entangled state can be expressed by \begin{equation}\label{psin}
           \psi_{1,n}=\int_{M}\psi(x_{1,n})|x_{1,n}\rangle dx_{1,n}\end{equation}as an integral over $M$. The differential $dx_{1,n}$ is short for $dx_{1}dx_{2},\cdots,dx_{n}.$ For each value of $x_{1,n},$ the integrand is a vector in $\bar{H}_{n}.$

           The state, $\psi_{1,n}$, is mapped onto the fiber bundle by treating the integrand, $\psi(x_{1,n})|x_{1,n}\rangle $ of Eq. \ref{psin} as a vector field, $\lambda_{x_{1,n}}$, that corresponds to a level $c$ section on the fiber bundle. In this case $\lambda_{x_{1,n}}$ becomes a vector in $\bar{H}^{c}_{x_{1,n}}$ in the fiber at $x_{1,n}.$ As was the case for  the two and one particle states, the integral of Eq. \ref{psin}, as an integral over the section, is not defined.  This is remedied by use of connections to map the integrands into a Hilbert space at a reference fiber location.  Let $v_{1,n}$ be the location. Here $v_{1,n}$ is called a location in $M$ even though it is an $n$ tuple of locations.

           The connection used for this case is an extension of that used in Eq. \ref{Cglamb12} for the two particle state. The action of the connection on the integrand at $x_{1,n}$, with $c$ replaced by $g_{v_{1,n}}$, is given by \begin{equation}\label{Cglamb1n}C_{g}(v_{1,n}:x_{1,n}) \lambda_{x_{1,n}} (u_{1,n})=\frac{g_{v_{1,n}}(z_{1,n})}{g_{v_{1,n}}(t_{1,n})}\lambda_{v_{1,n}}(z_{x,1,n}).\end{equation} In this expression $u_{1,n}$ is the tuple in $\mathbb{R}^{3}_{x_{1,n}}$  in the fiber at $x_{1,n}$ that is the chart equivalent of $x_{1,n}$ in $M$.  That is $\phi_{x_{1,n}}(x_{1,n})=u_{1,n}.$ Also $z_{1,n}$ and $t_{1,n}$ are tuples in $\mathbb{R}^{3}_{v_{1,n}}$ in the fiber at $v_{1,n}$ that are the $\phi_{v_{1,n}}$ chart equivalents of $x_{1,n}$ and $v_{1,n}.$ The subscripts on $\lambda$ and $g$ indicate the lifting of these quantities to local fields in the fibers whose location is indicated by the subscripts. For example, $g_{v_{1,n}}$ denotes the lifting of $g$ to a $\bar{C}^{g(v_{1,n})}_{v_{1,n}}$ valued scalar scaling field over $\mathbb{R}^{3}_{v_{1,n}}$ and $\lambda_{v_{1,n}}$ denotes the lifting of $\lambda$ to a $\bar{H}^{g(v_{1,n})}_{v_{1,n}}$ valued vector field over $\mathbb{R}^{3}_{v_{1,n}}.$

           The definition of $g_{v_{1,n}}$ is an extension of that for $g_{v,w}$ in Eq. \ref{g2vw}. One has
           \begin{equation}\label{gv1n}g_{v_{1,n}}(z_{1,n})=(\prod_{j=1}^{n}g_{v_{1,n}}(z_{j}))^{1/n}= \exp \frac{(\sum_{j=1}^{n} \gamma_{v_{1,n}}(z_{j}}{n}.\end{equation}Here $g_{v_{1,n}}(z_{j}$ and $\gamma_{v_{1,n}}(z_{j})$ denote the lifting of $g$ and $\gamma$ to
           $\bar{C}^{g_{n}(v_{1,n})}_{v_{1,n}}$ valued fields over $\mathbb{R}^{3}_{v_{1,n}}.$ Also $g_{n}(v_{1,n})$ is an extension to $n$ locations of the definition of $g_{2}$ in Eq. \ref{gxysq}.

           \section{Discussion}\label{D}

           The are many aspects of number scaling and the effects of scaling fields on quantum mechanics that should be discussed.  One is the use of a complex number value structure as separate from the different scaled representation structures. This is done to make clear the distinction between number and number value.  However  this may  not be necessary. The reason is that base set elements of representation structures automatically acquire values in any structure that satisfies the relevant axioms. Another distinct value structure should not be needed to describe this.

           Representations of base set elements of complex numbers that are  based on  rational number representations as lexicographically ordered symbol strings over an alphabet has been briefly described. More work is needed here  because these representations may provide a connection between experiment outputs as numbers and their values in different scaled representations.  This suggests that the value of the scaling field at the location at which an experiment or measurement is done will affect the value associated with the measurement outcome.  Note that a measurement outcome is a physical system in a specific state that is interpreted as a number. The value of the number may depend on  the scaling field value at the measurement location.

           The representations of quantum mechanical states of two or more  systems in fiber bundles with fibers base on two or more points in $M$ needs more work. One would like to have a single fiber bundle that includes quantum states of an arbitrary number of particles in each fiber. Here one needs separate bundles for $n$ and $m$ particle states.  One possible way to achieve this is to expand the treatment  to apply to relativistic quantum mechanics. Hopefully such an expansion would include quantum states in Fock space.

           The projection of the integrand for $1,2,\cdots, n$ particle states as a section on a fiber bundle has the result that the magnitude of contributions of the integrand for different $M$ locations depends on the locations and their separations. If the amplitude of the integrand as a vector field for a one particle state is very small at a location, then its contribution to the scaled wave packet at a reference location is very small.  This excludes the effect of the connection in moving the magnitude to the reference location.  The same argument holds for the dependence of the amplitude for two particle states on two reference locations and the distance between them.

           The choice of a reference location or locations for a fiber representation of the wave packet as an integral over the section is independent of the properties of the wave packet.  The location or locations can be anywhere with arbitrary separations between the locations. A Change of reference locations is a change of the location of the mathematical descriptions.  It is entirely separate from any physical change or translation of the quantum state.

           There is a rather philosophical reason for restricting reference locations to points in cosmological space and time that are accessible to us as intelligent observers.  For reference locations such as $v_{1,n}$ this  includes the collapsing of $v_{1,n}$ to one $M$ location such as $v_{j}=x$ for $j=1,2,\cdots,n.$ The reason is that mathematical structures, as models of axioms, have meaning or value. They are semantic structures.

           The important point is that the concept of meaning is observer related.  It is based on the idea that meaningfulness or value is a local concept associated with individual observers.  Mathematics in a textbook, described in a lecture, or on a computer screen, has no meaning or value until the information is transmitted by a physical medium, such as light or sound,  to the observers brain.  The concept of meaning or value is localized to the observers location. If the observer moves through space then the localization of meaning or value follows a path in $M$.  Since meaningfulness or value applies to many observers, there are many points in $M$ that are observer reference locations.

           As noted before, interest in local representations of physical and geometric quantities and the attendant scaling has its origin in gauge theories.  The  gauge freedom in these theories requires the use of separate vector spaces at each space time point. Unitary transformations between spaces at different locations represent the freedom of basis choice in each space. Since scalar fields, as real or complex numbers,  are part of the description of vector spaces, it seems natural to consider them also as localized. The freedom of choice of bases in vector spaces is expanded to include freedom of choice of scaling factors in the fields.

           Since numbers play such an essential role in physics it seems worthwhile to extend this localization and scaling factor choice to other mathematical structures used in other areas of physics. This includes those structures that include numbers as part of their description.  Examples include vector spaces, algebras, group representations as matrices, and many other systems. This is what is done here for complex numbers as used in quantum mechanics.

           A very important open question concerns the physical existence, if any, of the scalar scaling field, $g$. Candidates include the Higgs boson \cite{Higgs}, dark matter \cite{Saravani}, dark energy \cite{DE,Rinaldi}, inflaton \cite{I,Bezrukov}, quintessence \cite{QU}, etc. As noted elsewhere \cite{BenNOVA,BenINTECH}, the lack of direct physical evidence for the $g$ field means that the coupling constant of $g$ to fermion fields must be very small compared to the fine structure constant. This is required by the great accuracy of quantum electrodynamics without any $g$ field components.

           Some intriguing points are worth noting.  If the $g$ field has single particle excited states, then the particles would be scalar bosons, presumably of spin $0.$ The only  particle field with this property is the Higgs boson. Is the $g$ field related to the Higgs boson?

           Another point concerns the relation of dark energy to the physical vacuum energy as expanding space time. The effect of scaling on geometric quantities as in \cite{BenNOVA,BenQSMF} is in some ways similar.  As was noted, the number $0$ is unique as the only number whose value is independent of scaling.  In this sense it is like a number vacuum. Assume that the scaling factor is time dependent.  Then the distance between points in $M$, as a number value in $\bar{R}^{g(x,t)},$ is a function of $x$ and $t$.  If $d$ is a  distance of two points at time $t_{0},$ then the distance between the same points at time $t_{1}$ is $(g(t_{1},x)/g(t_{0},x))d$ \cite{BenQSMF}. Is this related to dark energy?  Answers to these and other questions is work for the future.

           The lack of physical evidence for the effect of the $g$ field in quantum mechanics, as shown here, means that the variation of the $g$ field over the region of cosmological space and time in which experiments have been conducted is below experimental error.  This is equivalent to the requirement that the vector field $\Gamma(x),$ as the gradient of $\gamma(x)$ be so small as to have avoided detection in the region of experiments done to date.   This is the region occupied by us as beings capable of carrying out experiments and making theory predictions.

           It is impossible to predict what the future will say about the $g$ field or its gradient.  However it is the case that  all experiments done by us, in the past,  now, or in the future, will be done in a region of cosmological space and time that is either  occupiable by us, or by other intelligent beings on distant planets with whom we can communicate effectively. One estimate \cite{AmSci} of the size of this region is as a sphere of about $1200$ light years in diameter that includes the solar system. The exact size of the region is not important.  The only requirement is that it is small compared to the size of the universe.

           The result of these considerations is that there are no restrictions on the properties of the $g$ field in  space and time regions outside  of the occupiable one. Nothing prevents the $g$ field from being very large with rapidly varying gradients in these regions.  This applies to regions close to the big bang, about $14$ billion years ago, as well as to most of the universe.

           It is clear that there is much more work needed.  This includes expansion of the manifold to space time to include special relativity and  to a pseudo Reimannian manifold for general relativity. One may hope that these expansions will offer some clues to the physical nature of the scalar scaling field.

             \section*{Acknowledgement}
            This material is based upon work supported by the U.S. Department of Energy, Office of Science, Office of Nuclear Physics, under contract number DE-AC02-06CH11357.


\begin{thebibliography}{99}

            \bibitem{Wigner}
            E.  Wigner,  "The unreasonable effectiveness of mathematics in the natural sciences", Commun.  Pure Appl. Math. \textbf{13}, No. 1, (1960).  Reprinted in E. Wigner,
           \emph{Symmetries and Reflections}, (Indiana Univ. Press, Bloomington, IN 1966), pp222-237.

             \bibitem{Omnes}
            R. Omnes,   "Wigner's "Unreasonable Effectiveness of
           Mathematics", Revisited," Foundations of Physics,
           \textbf{41}, 1729-1739, (2011).

            \bibitem{Plotnitsky}
            A. Plotnitsky, "On the reasonable and unreasonable effectiveness of mathematics in classical and quantum physics" Found. Phys. \textbf{41}, pp. 466-491, (2011).

              \bibitem{Tegmark}
            M. Tegmark, "The mathematical universe", Found. Phys., \textbf{38}, 101-150, 2008.

              \bibitem{BenCTPM}
            P. Benioff, "Towards a coherent theory of physics and mathematics" Found. Phys. \textbf{32}, 989-1029, (2002).

            \bibitem{BenCTPM2}
            P. Benioff, "Towards a coherent theory of physics and mathematics: the theory-experiment connection",  Found. Phys., \textbf{35}, 1825-1856, (2005).

            \bibitem{BenNOVA}
             P. Benioff,  "Gauge theory extension to include number scaling by boson field: Effects on some aspects of physics and geometry," in \emph{Recent Developments in Bosons Research}, I. Tremblay, Ed., Nova publishing Co., (2013), Chapter 3;  arXiv:1211.3381.

            \bibitem{BenQSMF}
            P. Benioff, "Fiber bundle description of number scaling in gauge theory and geometry", Quantum Stud: Math. Found. \textbf{2}, 289-313, (2015), arXiv:1412.1493


                \bibitem{Yang}
            C. N. Yang and  R. L. Mills,  "Conservation of Isotopic Spin and Isotopic Gauge Invariance," Phys. Rev., \textbf{96}, 191-195,  (1954).

             \bibitem{Shapiro}
             Shapiro, S.: Mathematical Objects, in Proof and other dilemmas, Mathematics and philosophy, Gold, B. and  Simons, R., (eds), Spectrum Series, Mathematical Association of America, Washington DC, 2008, Chapter III, pp 157-178.


             \bibitem{Barwise}
           J. Barwise,  "An Introduction to First Order Logic," in
            \emph{Handbook of Mathematical Logic}, J. Barwise, Ed.
            North-Holland Publishing Co. New York, 1977. pp 5-46.

            \bibitem{Keisler}
            H. J. Keisler, "Fundamentals of Model Theory", in
            \emph{Handbook of Mathematical Logic}, J. Barwise, Ed.
            North-Holland Publishing Co. New York, (1977). pp 47-104.

             \bibitem{Czachor}
             M. Czachor, "Relativity of arithmetics as a fundamental symmetry of physics",
              arXiv:1412.8583

            \bibitem{Husemoller}
             D. Husem\"{o}ller, \emph{Fibre Bundles}, Second edition,  Graduate texts in Mathematics, v. 20, Springer Verlag, New York, (1975).

             \bibitem{Husemoller2}
              D. Husem\"{o}ller, M. Joachim, B. Jurco,  and M. Schottenloher, \emph{Basic Bundle Theory
              and K-Cohomology Invariants},  Lecture Notes in Physics, 726, Springer, Berlin, Heidelberg, (2008), DOI 10.1007/ 978-3-540-74956-1, e-book.



              \bibitem{Daniel}
              M. Daniel and G. Vialet, "The geometrical setting of gauge theories of the Yang Mills type", Reviews of Modern Phys., \textbf{52}, pp 175-197, (1980).

              \bibitem{Drechsler}
              W. Drechsler and M. Mayer, \emph{Fiber bundle techniques in gauge theories}, Springer Lecture notes in Physics \#67, Springer Verlag, Berlin, (1977).

              \bibitem{Moylan}
              P. Moylan, "Fiber bundles in nonrelativistic quantum mechanics", Fortschr. Phys. \textbf{28}, pp 269-284, (1980).

              \bibitem{Sen}
              R. Sen and G. Sewell, \emph{Fiber bundles in quantum physics}, Jour. Math. Physics, \textbf{43}, 1323-1339, (2002).

              \bibitem{Bernstein}
              H. Bernstein and A. Phillips, \emph{Fiber bundles and quantum theory}, Scientific American, July, (1981).

              \bibitem{Iliev}
              B. Iliev, \emph{Fiber bundle formulation of nonrelativistic quantum mechanics}, arXiv:quant-ph/0004041.

              \bibitem{Asorey}
                M. Asorey, J. Cari~nena, M. Paramio, "Quantum evolution as a parallel transport",
                J. Math. Phys. \textbf{23}(8), 1451-1458, (1982).


             \bibitem{BenINTECH}
             P. Benioff, "Effects on quantum physics of the local availability of mathematics and space time dependent scaling factors for number systems", in \emph{Quantum Theory}, I. Cotaescu, Ed.,  Intech open access publisher, 2012, Chapter 2, arXiv:1110.1388.


            \bibitem{complex}
           J.  Shoenfield, \emph{Mathematical Logic}, Addison Weseley
            Publishing Co. Inc. Reading Ma, (1967), p. 86; Wikipedia:
            Complex Numbers.

             \bibitem{Cauchy}
            E. Hewitt and K. Stromberg,  \emph{Real and Abstract Analysis}, Springer-Verlag New York,  Inc. (1965), Chap. I, Sect. 5.


              \bibitem{BenSPIE5}
              P. Benioff, "Principal fiber bundle description of number scaling for scalars and vectors: Application to gauge theory",{Quantum Information and Computation XIII}, E. Donkor,  A. Pirich, and M. Hayduk, Eds., Proceedings of SPIE, Vol. 9500; SPIE: Bellingham, WA, 2015, 98227, arXiv:1503.05600.

            \bibitem{Montvay}
            I. Montvay and G. M\"{u}nster, \emph{Quantum fields on a lattice}, Cambridge University Press, UK,(1994), Chapter 3.


            \bibitem{Higgs}
              P. W. Higgs,    "Broken symmetries and the masses of gauge bosons".
             Phys. Rev. Lett. 13 (16): 508, 1964.

             \bibitem{I}
             A. Albrecht and P. Steinhardt, "Cosmology for grand unified theories with radiatively induced symmetry breaking", Phys. Rev. Letters, \textbf{48},1220-1223, (1982).

             \bibitem{Saravani}
             M. Saravani and S. Aslanbeigi, "Dark matter from spacetime nonlocality", arXiv:1502.01655.


             \bibitem{DE}
             M. Li, X-D. Li, S. Wang, Y, Wang, "Dark energy, A brief review", arXiv:1209.0922.

             \bibitem{Rinaldi}
             M. Rinaldi, "Higgs dark energy", arXiv:1404.0532v4.

             \bibitem{Bezrukov}
             F. Bezrukov and M. Shaposhnikov,  "The standard model Higgs boson as the inflaton",  Physics Letters B, \textbf{659},703-706, (2008).

             \bibitem{QU}
              I. Zlatev, L. Wang, L. P. Steinhardt,   "Quintessence, Cosmic Coincidence, and the Cosmological Constant". Physical Review Letters 82 (5): 896–899, 1999; arXiv:astro-ph/9807002.

             \bibitem{AmSci}
           H. A. Smith, "Alone in the Universe", American Scientist, \textbf{99},  No. 4, p. 320, (2011).




            \end{thebibliography}
            \end{document}